\documentstyle[12pt,aasms4]{article}

\def\etal{{\it et al.}}

\def\putplot#1#2#3#4#5#6#7{\begin{centering} \leavevmode
\vbox to#2{\rule{0pt}{#2}}
\includegraphics{#1}

\end{centering}}

\begin{document}


\title{Late-type dwarf irregular galaxies in the
Virgo cluster. \\ I. H$\alpha$ and red continuum data}


\author{Ana Heller and Elchanan Almoznino }

\affil{The Wise Observatory and 
the School of Physics and Astronomy \\ Tel Aviv University, Tel Aviv 69978,  
Israel}

\and
 
 \author{Noah  Brosch\altaffilmark{1}}
\affil{Space Telescope Science Institute \\ 3700 San Martin Drive \\
Baltimore MD 21218, U.S.A.}

\altaffiltext{1}{On sabbatical leave from the Wise Observatory and 
the School of Physics and Astronomy,
Raymond and Beverly Sackler Faculty of Exact Sciences,
Tel Aviv University, Tel Aviv 69978, Israel}


\begin{abstract}
 
We present H$\alpha$ and red continuum observations for a 
sample of late-type low surface
brightness (LSB) dwarf irregular galaxies, consisting of
all the ImIV and V galaxies with m$_B\leq$17.2 in the Virgo cluster,
 and compare them with similar data for a representative
sample of high surface brightness (HSB) dwarf irregular galaxies, 
also in the Virgo cluster. Line
fluxes and equivalent widths are listed for individual HII
regions and total H$\alpha$ emission is measured for the entire
galaxies. Although significant line emission originates in
the HII regions we identified, it does not make up the entire 
 H$\alpha$ output of all galaxies. 

 For those objects of the LSB sample with H$\alpha$ 
emission we find typical star formation rates (SFR) of 6.9 10$^{-3}$ 
M$_{\odot}$ yr$^{-1}$, to 
as high as 4.3 10$^{-2}$ M$_{\odot}$ yr$^{-1}$. This is,
on average, one order of magnitude weaker than for HSB objects,
although the SFR overlap. On
average, $\sim$2 HII regions are detected per LSB galaxy, for a total of
38 HII regions among 17 galaxies with H$\alpha$ emission. The HII regions 
are smaller and fainter than in HSB galaxies in the 
same Virgo cluster environment, have H$\alpha$ line equivalent widths
about 50\% of those in HSBs, and cover similar fractions of the galaxies.
When more than one HII region is present in a galaxy we observe a
strong intensity difference between the brightest and the second
brightest HII region. The line-emitting regions of LSB galaxies
are preferentially located at the periphery of a galaxy, while
in HSBs they tend to be central. The H$\alpha$ line strength
of an HII region is correlated with the red continuum light underneath
the region; this holds for both LSBs and HSBs.

We do not identify fundamental differences in the star formation
properties of the LSB and HSB dwarf galaxies we studied and infer that 
these galaxies must be similar, with the
difference being the intensity of the present star-formation
burst.

\end{abstract}
\vspace{1cm}


\vspace{1.2 truecm}
\section{Introduction}

The surface brightness of a galaxy is primarily determined by 
 the stellar luminosity function and by the projected  density of stars
and these depend on the past and ongoing star formation.
Much of our current understanding  of the star formation (SF) phenomena
on all scales is biased by the high surface brightness (HSB)
objects that define the classical Hubble sequence.  However, irregular
galaxies in general, and dwarfs in particular, offer unique tests
of SF models since they lack the more complex SF-triggering 
mechanisms, such as prominent spiral arms with the associated 
density waves and differentially rotating disks.

Low surface  brightness (LSB) 
galaxies are defined as objects with  central blue surface
 brightnesses   $\mu(B)_o >  23$ mag  arcsec$^{-2}$.
Recent studies of LSB spiral (Sc and later) galaxies
(McGaugh {\it et al.}~1995; de Blok {\it et al.}~1995) indicate that
  these are slowly evolving systems.  
The indication of a slow evolution but not full quiescence 
is based on the blue colours,  presence of HII regions, low metallicites, 
extended gas disks, and high M(HI)/L ratios. The relative
isolation of these systems and the low density of their gas component may
be the reasons for the small amplitude SF bursts, and therefore the
low overall star formation rates (SFR).  
  
The SF history of a quiescent galaxy with a
   suitable  gas reservoir in a cluster environment  may be  influenced
 by external factors such as the temperature of the hot 
intra-cluster medium (ICM) and the galaxy's interaction with other
cluster galaxies and with the intra-cluster medium. Therefore it is
important to choose carefully the environment of a sample of galaxies
studied to learn about star formation processes. Recent work on  HSB 
 dwarf galaxies (DGs) in the Virgo cluster (blue compact dwarf
BCD or ImIII/BCD: 
Almoznino~1996; Almoznino \& Brosch 1998, AB98) points to a possible 
evolutionary connection between their
rather blue underlying stellar component  and  LSB DGs.
 We  approach these questions here by  analyzing optical narrow and 
broad-band CCD observations of fainter luminosity objects (ImIV-V) 
in the same environment, the Virgo cluster (VC).
 Both samples (HSB and LSB) comprise altogether 47 late-type irregular 
galaxies.

In this paper we concentrate on the morphology of HII regions in
our sample of LSB galaxies, and on the star formation rates (SFR)  derived 
from galaxy-wide H$\alpha$ fluxes as well as from individual HII regions
for the objects in the LSB sample. We compare these with similar
observables from the HSB sample of dwarf galaxies presented in AB98.
The  H$\alpha$ radiation is directly coupled to the radiation shorter than 
$\lambda < 912$\AA\, and originates from OB stars of lifetime
$ \simeq 5 \times 10^{6}$ years, indicating on-going massive star formation,
while optical broad-band colours provide information on older stellar populations.
 The surface photometry of the galaxies in our sample through UBVRI filters,
with calibrated  fluxes, radial
 surface brightness, contour colour maps and profiles, central surface 
brightness, and scale lengths shall be presented in forthcoming papers
along with a full analysis of the entire data set.
 
The plan of the paper is as follows: we first describe the sample
in section 2, then the observations we performed and their reduction
in section 3. The results are discussed in section 4 and our findings
are summarized in section 5.

 \section{The sample}

The selection of LSB galaxies is based primarily on the morphological
classification of DGs in the VC  (Binggeli {\it et al.}~ 1985, VCC).
We selected galaxies listed in VCC as Magellanic Irregulars
of sub-types ImIV or ImV with $m_{B}\leq17.2$ and members
in the VC. The faintest ImV galaxies listed in VCC have $m_{B}=17.5$, 
thus our selection is safely distant from  the limit of the VCC cutoff. 
All our  galaxies have single dish HI
 measurements with detected non-zero total flux from the Arecibo survey of  
 VCC late-type dwarf galaxies of Hoffman~{\it et al.}~(1987). 
We imposed a recession velocity limit of 3,000 km sec$^{-1}$
to prevent the inclusion of background objects.
 
We adopt here a uniform sample distance of 18 Mpc for which 1"=87 pc. This 
assumption is
justified as being the representative distance to galaxies within
the Virgo cluster ({\it e.g.,} Fouqu\'{e} \etal \, 1990). Other
distance estimates are in the same range: 16.0$\pm$1.5 Mpc (van den 
Bergh 1996), 17.2-17.4 Mpc (Gonzalez \& Faber 1997), 20.7$\pm$2.4 Mpc
(Federspiel \etal \, 1998). Note though that
the VC has significant depth, thus the assumption that all galaxies
are at the same distance is probably true only at the 10\% level. As most of 
our conclusions are drawn from ratios of quantities measured in
the same galaxy, an even larger error in the distance will have no effect.
The  fluxes and flux densities would, however, change.

The sample, as defined here, is complete and is representative of 
LSB dwarf galaxies, though not of the lowest surface brightness ones.
The reason is that galaxies with even fainter surface brightness
exist, in the same neighborhood, and to study them as done here for 
ImIV-V objects is difficult and requires extensive observing time 
at major telescopes.
 
The 27 galaxies in the LSB sample are listed in Table 1. We identify them
by their number in the VCC and list their total blue magnitude
(which is an eye estimate from VCC), the total red magnitude (measured 
here, as explained below,
and presented with a measurement error), 
the measured semi-major
axis {\it a} of the optical image in arcsec and ellipticity 
$e=1-\frac{b}{a}$ derived
from an ellipse fit to the 25 mag arcsec$^{-2}$ isophote, 
and the H$\alpha$ results. The latter are given as total flux,  
total equivalent width, and fractional coverage of a galaxy by
HII regions. The projected distribution of the objects in the VC 
is shown in Figure 1, where each object is marked as a circle with
a diameter proportional to its apparent magnitude. 

We added in the figure the galaxies from the HSB sample (filled circles,
N=16) discussed in 
AB98, and indicated the various sub-clusterings in the VC identified in
Hoffman \etal \, (1989). The HSB sample shall be compared at various 
points of this paper with the LSB objects and some information, appearing 
already in AB98, shall be repeated here. This HSB comparison sample is
composed of irregular galaxies, and the objects were selected to belong
to the morphological class ``blue compact dwarf'' (BCD).

\section{Observations and Reduction}

Most LSB images analyzed here were obtained with the Wise Observatory 
(WO) 1.0-m telescope
using a TEK 1024$\times$1024 pixel thinned and back-illuminated CCD. 
This chip has a very low read-out noise (6.5 e$^{-1}$) and its
quantum efficiency in the regions of interest is high ($\sim$80\% at
6500\AA\,). The plate scale for the WO images is 0".7 pixel$^{-1}$. 
One galaxy studied here was observed by EA in 1992 at the WO using 
an RCA CCD  with 0".9 pixels (see AB98). A few  
exposures were obtained by NB at the $f/4$ Prime Focus of the Russian Academy 
of Sciences' Special Astrophysical Observatory 6.0-m (BTA) telescope. 
The CCD used for those observations was 
a 512$\times$512 array, with a projected pixel size of  0.154"$\times$0.205".
The readout noise of the device used on the BTA was 30 e$^-$ and the 
conversion from DNs was 13.8 e$^-$=1 DN. The BTA images were binned
off-line $4\times3$ pixels, which gives 0".615 square pixels, comparable 
to the plate scale of the WO images.

The image sets analyzed here consist of line and continuum images for 
each galaxy obtained from 1995 to 1997. We collected three 1800 sec (WO), 
or two 600 sec (BTA), exposures through the line filter and an equal 
number of continuum
exposures per object. The line filters were chosen to match, as far 
as possible, the rest frame H$\alpha$ emission of each galaxy and are
described in AB98. 
We used a $\sim$53\AA\, filter centered on 6700\AA\, for measuring 
the continuum contribution.
 The reduction was done with IRAF and the flux calibration used 
observations of HZ44, which was measured several times each night through 
different air masses with the same filters as used for the galaxies.  
Note that though this standard star shows H$\alpha$ and HeI absorptions,
these do not affect the calibration because the filters used here
are redshifted with respect to these features in the standard.

The images were debiased, flat fielded with twilight frames,
 and cosmic rays and hot pixels were removed with  
standard IRAF tasks. The images through the same filter were 
 co-added  after being registered by cross-correlation.
The sky background was subtracted as a tilted plane
 fitted to the final combined
image in each band, after the galaxy and the bright stars were masked off. 
The standard deviation of the fitted plane  from the actual background 
light distribution was used  to estimate the
noise introduced by sky subtraction. 

The images were calibrated in flux using the observations of  HZ44
and were convolved with a 2"-4" FWHM gaussian, according to the seeing. 
We then performed aperture photometry on stars in the field, 
normalizing the results of the line and continuum bands. In this way we 
 also corrected for any  relative change 
of continuum sloping in the two bands. The scaling
between the on-line to off-band images was applied to the
full resolution images and the net  fluxed H$\alpha$
images were derived by subtracting off-line from on-line fluxed images.
 As expected, the stars vanished from the net image, with the 
exception of cases where imperfect PSF matching caused  small 
distortion near very bright stars. This method is very similar to that
used successfully by {\it e.g.,} Hodge (1969), Hunter \& Gallagher (1986), and
recently Gavazzi \etal \, (1998).

The photometry, and the derivation of line and continuum fluxes, are
based on measurements of the HZ 44 spectrophotometric standard and flux
scaling from it to the galaxy, after correcting for differential 
extinction. We estimate the photometric accuracy from the extinction
coefficients obtained for the ON and OFF bands; these are both 0.19$\pm$0.12
and are consistent with the long-term average of the R-band extinction 
coefficient at the Wise Observatory (0.21; Brosch 1992). 
As the typical airmass differences between observations of the standard 
and of the target galaxies were always $\leq$0.3, this indicates that
the error which could be attributed to extinction is at most $\sim$0.06 mag.

 The off-line continuum images were smoothed to a 1" resolution and  
iso-intensity contours were  traced  around each galaxy.  
Polygonal apertures were defined by the contour level of
25 mag arcsec$^{-2}$   isophote. Generally, this 
isophote level corresponded to a $S/N=1$ per pixel in the final
images for all galaxies.   
The total H$\alpha$ flux from the galaxies was obtained by 
measuring within the polygonal aperture the averaged $f_{\lambda}$  in the 
net emission images and multiplying the results by the FWHM of the 
H$\alpha$ line filter. We did not compensate for the
 contribution of [NII] lines to the derived 
H$\alpha$  fluxes, as the ratio $\frac{[NII]} {H\alpha}$ in 
dwarf irregular galaxies is expected 
to be small  due to their generally low metal 
abundances (Gallagher \& Hunter 1989, Skillman  \etal \, 1989).
 
The red continuum emission from each galaxy, f(OFF), in flux
density units, was also measured within the same polygonal
aperture. The total flux was transformed into monochromatic
magnitudes at $\sim$6700\AA\, (m$_{67}$), which are
listed in Table 1. The ratio between the net H$\alpha$ emission and
the red continuum flux measured from the off-line image is the emission 
line equivalent width (EW). Although 
it is not yet clear which stellar population is responsible for 
this red emission, {\it i.e.,} whether it is produced by the
stars formed in the present SF event or by a pre-existing
stellar population, we list these EW values along with the
line fluxes in Table 1. The
``monochromatic'' magnitude at $\lambda\simeq$6700\AA\, is not equivalent
to the R-band magnitude;
a transformation of the flux density for a zeroth magnitude star in Bessel
(1979) for the R$_c$ band into the monochromatic magnitude system yields
m$_{67}\simeq$0.44. Therefore, the use of this system introduces
some zero-point shifts. The errors, in  both the line and 
continuum measurements, are 
derived from the rms combination of the errors of the polygon
photometry of the galaxy and the calibration of the spectrophotometric
standard star.

We could not measure the dimensions of the red continuum image for VCC
381 because of its exceeding faintness. The numbers quoted in parentheses 
in Table 1 are from NED. In the
case of VCC 2034 a star is projected against the galaxy; we could not
subtract accurately its contribution to enable a reliable determination
of m$_{67}$. Similar data for the HSB sample are presented in 
Table 2. This is done here
only for the purpose of completeness and comparison; the full data set
relevant for the HSB galaxies is part of the AB98 paper and is discussed 
there. We note that the
entire sample consists of ``dwarf'' galaxies, as they all have 
M$_B>-18$. The two samples studied here are thus representative of the
late-type dwarf irregular kind, and they come in two ``flavors''
of high or low surface brightness.

Table 2 lists the m$_B$ and m$_R$ magnitudes for the HSB galaxies;
these are the values actually measured through the broad-band B and
R filters (where the latter contains H$\alpha$ and [NII]
contributions). The level of line contamination of the R magnitude
can be estimated from the equivalent width of H$\alpha$. In
contrast, Table 1 lists the m$_B$ estimated in the
VCC for the LSBs, and gives as m$_{67}$ the measured monochromatic magnitude
of the continuum, which is free of line contamination.

Table 3 presents the statistics of detections of HII regions in our
two samples (HSB from AB98 and LSB discussed here) and the last
column of Tables 1 \& 2 lists the on-going global average  SFR of 
each galaxy, derived from
the relation of Kennicutt \etal \, (1994) scaled to the 18 Mpc adopted
distance of the VC: SFR=2.93 10$^{11}$ F(H$\alpha$), where F(H$\alpha$) 
is the total line flux  in erg s$^{-1}$ cm$^{-2}$ and the SFR is 
in M$_{\odot}$ yr$^{-1}$.

Finally, we also measured the fluxes from the individual HII regions
which were identified in the H$\alpha$ images by integrating the
net flux collected within  a circular aperture around each HII region. 
The size of this aperture was determined by first visually inspecting the
net line image using the IMEXAM task from IRAF, followed by PHOT with a
number of apertures. This aperture size taken here to represent the 
size of the HII region is that where an increase in size did not add
more flux. The individual HII
region fluxes are listed in Table 4. We also list there the SFR derived 
for each  HII region, as well as the flux in the red continuum underneath it,
along with the line EW and the individual SFR derived for the specific 
HII region, as explained above. The
errors listed for the HII regions are only photometric. For completeness, 
we list in Table 5 similar parameters for the HII regions in the HSB sample. 
The HII regions are identified by the
name of the galaxy in column 1, and by the label given to the HII region in
the specific figure in column 2. The HSB data equivalent to these presented 
in Table 1 for the LSB galaxies have been published in AB98.

\section{Results }

Figures 2 through 7 present mosaic images of the LSB galaxies observed at
the WO. We show three to five galaxies per figure in horizontal rows. Each
row contains the image through the line filter at the left, the
image through the continuum filter in the middle, and the net H$\alpha$
image at the right. Each top-right image on a page of LSBs has a
scale mark near its top-right corner; the mark is 10 arcsec long
(=870 pc at the adopted distance of the VC)
and indicates the scale of all the images on the page. Fig. 
8 shows the two galaxies observed at the BTA and the single galaxy from the
LSB sample observed at WO by EA with a different CCD.
Figs.  9, 10, and 11 show the BCD galaxies from AB98 in the same
format as used here for the LSB objects, for comparison and completeness.
These images have not been presented elsewhere.
Figs. 8 to 11 have individual 10 arcsec bars for each object.
  For each galaxy we marked and labelled the individual HII regions we
recognized in the net H$\alpha$ image.

The underlying LSB galaxies, measured on the red continuum images, have
an average major axis diameter of 58"$\approx$5 kpc. For comparison, the 
average major axis of the HSB sample is 40"$\approx$3.5 kpc.The semi-major 
axes of the LSB galaxies are listed in Table 1, along
with the ellipticity parameter $e$. These were derived from the fit 
of an ellipse to the red continuum light at the 25 mag arcsec$^{-2}$
isophote. We also list there $f$, the fractional coverage of a galaxy by
HII regions. This is obtained by dividing the sum of all aperture areas with
which we measured the individual HII regions by the area of the
``continuum'' galaxy, defined by the measuring polygon described above.

We identified 38 HII regions, 
single or in complexes, in 17 galaxies of the LSB sample.  
Due to the limiting spatial resolution of this 
study ($\sim$300 pc at the VC), and the limiting surface brightness, 
we were not able in some cases to separate individual small HII regions
from their nearest neighbours, nor to detect the faintest HII regions.
Therefore our survey was sensitive to HII ``complexes'' of typical 
sizes larger than the 300 pc resolution limit or to giant HII regions.
About half of the galaxies with H$\alpha$ emission have a single HII
region and the rest have two or more regions. In two cases (VCC 17
and VCC 826), and possibly in VCC 328 as well, the HII regions seem to be
arranged on the circumference of a circle or an ellipse. In VCC 17 
it seems that the center of this distribution
shows an HII region as well; in VCC 826 the center seems empty.
We emphasize that the localization of the HII regions on the circumference
of an ellipse is a subjective result; we have not attempted to fit such
an ellipse to the distribution of the HII regions, and indeed such a fit may
not succeed in a number of cases because of the strong intensity
 of one HII region compared with others in the same galaxy.

In contrast with the LSB galaxies, the HSB objects show, in most cases, 
strong H$\alpha$ emission originating from a single HII region. The
line-emitting region is similar in size to the entire galaxy. 
The global H$\alpha$ emission, in terms of the line
equivalent widths,  shows that the LSB sample has more low EW
values and the global line emission hardly reaches 100\AA\,, 
whereas in the HSB sample $\sim$1/3 of the HII regions reach or exceed this
value. The entire issue of the
morphology of star formation in irregular galaxies is discussed
in another paper (Brosch \etal \, 1998a).

The data in Table 3 indicates that the difference between HSB and LSB galaxies 
is more pronounced when considering HII regions than for entire galaxies. The difference
in median SFRs for individual SF regions is more than one order of magnitude;
the HSBs have much more intense SF than LSBs. 
We compare in Fig. 12 the SFRs found for individual HII regions in the 
 sample of HSB galaxies
from AB98 and in the sample of LSBs studied here. We included all the
listed H$\alpha$ values, not only those higher than the 1.5$\sigma$ level. 
 In general, HII regions in HSB galaxies 
show more intense SFR, by 0.5 dex on average relative to HII 
regions in LSBs. However, there are HII regions in objects
classified as Blue Compact Galaxies which show  SFR levels as low
as those in some LSBs. This is not exceptional, as in
both groups of galaxies there are cases where no H$\alpha$ emission
is observed. In addition, we note that not only is the H$\alpha$ 
emission from HII regions in HSBs more intense than in LSBs, but also the
line equivalent width is higher (median H$\alpha$ EW in LSBs is 
47\AA\, {\it vs.} 80\AA\, in HSBs).

\section{Discussion}

The properties of star formation, derived from the surface brightness 
of H$\alpha$, have been discussed by Brosch \etal \, (1998b).
Briefly, in dwarf irregulars the average H$\alpha$ surface brightness 
correlates reasonably well with the mean blue surface brightness ($\sigma_B$). 
If the SF process is regulated on a local scale, the correlation
should be tighter if we consider only the HII regions, and not the
entire galaxies. We therefore test here the correlation of H$\alpha$ 
emission against the red 
continuum surface brightness for individual HII regions. This
is preferred to $\sigma_B$, because it is determined here independently
as part of the derivation of the H$\alpha$ brightness. We show in
Figure 13 the correlation between the H$\alpha$ flux of an HII region
and the red continuum emission underlying the region, as measured 
on the off-line image with the same aperture as used for the 
H$\alpha$ measurement of the LSB sample (stars). Although a
trend for more intense line emission when the red continuum is more intense
seems to be present, it is by no means tight. A formal linear
regression indicates a correlation coefficient of 0.62 (F=23)\footnote{F is 
the ratio between the mean square deviation due to the regression and
the mean square deviation due to the residual variation. For a linear 
regression, which is the present situation,
F=t$^2$ and this is the equivalent of a t-test. For more details
see Draper \& Smith 1981.}. A 
similar regression for the fewer HII regions in BCDs (filled squares) 
shows a tighter relation (correlation coefficient 0.82, F=32).
Both correlations are significant and indicate the regulation of
the SF activity by the local population of stars. In particular,
note that the two samples ``line-up'' nicely on the plot, indicating
that the connection between H$\alpha$ line strength and underlying
line continuum brightness transcends the morphological type of
a galaxy.

An inspection of Table 4 and of
Figs. 2 to 7 indicates that in most cases of multiple HII regions 
in LSB galaxies one is much brighter than the
others. This indicates that either the star formation does not occur 
simultaneously within the galaxy, or that if it does, not all star-forming
sites reach the same strength of star formation at the same time. 
The fraction of objects
with more than one HII region among the HSB sample is much smaller,
but the cases of VCC 1374, and to a lesser degree of VCC 1725 and
VCC 1791, are similar. We can quantify this difference between HII regions in
the same galaxy with the data in Table 4: the H$\alpha$ flux ratio between 
the brightest
HII region and the second brightest is $\sim$3 (median 2.3) and the similar
ratio of equivalent widths is $\sim$2. 

Not all H$\alpha$ emission from an irregular galaxy originates in the
bright HII regions. We compared the summed H$\alpha$ line flux from the HII
regions of one galaxy listed in Tables 3 and 4 with the integrated 
line emission from the same galaxy (Tables 1 and 2)  and
found that a significant fraction of the H$\alpha$ flux is not recovered. 
In particular, for 
VCC 17, VCC 260, and VCC 530 the HII regions we identified produced 
less than 70\% of the total galaxy line output (but more than 50\%).
In a number of objects where we did not detect individual HII regions
we nevertheless measured a significant H$\alpha$ flux (VCC 329, VCC
1816, and VCC 2034; all detected in the line at the 3$\sigma$ level
and all from the LSB sample). However, in eight objects it seems that 
virtually the entire H$\alpha$
output originates in the individual HII regions. It is clear that the detection
of an HII region is a combination of compactness, signal strength,
and contrast against the background of this region.

We find that the fractional coverage of a galaxy by HII regions (the ratio
between the summed areas of the apertures with which we measured
the HII regions and the total area of a galaxy on the red continuum image)
is amazingly similar between LSB and HSB objects and in both 
cases it peaks near 50\%, as Figure 14 shows.  The average value
of $f$ in HSB objects is $\sim2\times$ larger than in LSB objects, but 
the median values are similar: 38\% for the LSBs and 43\% for the HSBs.
As is often the case with small samples, the median values are more
representative of the ``typical'' behavior than the means, which are 
driven by outlier values. However, it seems that the HSB galaxies
lack the faint HII regions which are detected in the LSB sample.

The location of HII regions within the red continuum envelope, which
presumably traces the distribution of old stars or of red supergiants,
shows an interesting trend: in the large majority of cases where a
galaxy has HII regions, these appear preferentially at the periphery
of the red envelope. Only two cases of possibly central (``nuclear'')
HII regions have been identified: VCC 963 and VCC 1455. In three
other cases the HII regions were indiscriminately distributed over the
red surface of the galaxies, with no special preference to either
center or periphery. 

The trend  of SF activity to be relegated
to the periphery of irregular galaxies is discussed in more
detail in Brosch \etal \, (1998a) and could, in principle, indicate a
coupling between the interstellar matter in a galaxy 
 and the gas in the surroundings of the galaxy. This has been suggested
as a possible explanation of lop-sided SF activity in the LMC
(de Boer \etal \, 1998). In
this case, the SF activity could be triggered by some sort of shear or
pressure which compresses the ISM in the galaxy and starts making stars, 
which then move away from the location of formation and age. The difference 
between the brightest HII region and the other HII regions of the same LSB 
galaxy could, in this case, be the result of aging. The prediction from this
scenario is that
as an HII region gets older its total line emission flux decreases 
while the underlying red continuum emission increases (because of the
contribution from the aging stars on top of that from any
previous stellar populations), thus its line EW decreases sharply. 
We can test this using the data presented here.

We calculated the ``contrast'' in line emission and in underlying continuum
emission for those objects which show multiple HII regions. Specifically,
we calculated the ratio between the H$\alpha$ fluxes of the most intense
 HII region and the least intense one in a galaxy, and plotted this
in Figure 15 against the ratio of the red continua flux densities of the same two
HII regions. The distribution of points shows that the stronger H$\alpha$
emission is related to the more intense red continuum, an opposite trend 
(to first order) than expected from an assumption of SF
activation by interaction with external gas. The correlation is low
(0.51, F=4) but indicative and the trend, being opposite to what was
expected, allows us to reject the hypothesis. On the other hand,
a regulation of the SF by the local star density would make a plausible
possibility and would fit well with our finding that the SFR per unit
solid angle correlates with the blue surface brightness (Brosch
\etal \, 1998b).

\section{Conclusions}

We studied a sample of very low surface brightness irregular galaxies
in the Virgo cluster and found that the majority show indications of
current star formation. We summarize our findings as follows:

\begin{enumerate}

\item The LSB galaxies have a modest total SFR, typically of 
$\sim$7 10$^{-3}$ M$_{\odot}$ yr$^{-1}$, while HSB galaxies have
a typical SFR of $\sim$6 10$^{-2}$ M$_{\odot}$ yr$^{-1}$.

\item The typical global H$\alpha$ equivalent width of LSBs is 
$\sim2\times$ lower than in BCDs. 

\item The HII regions of BCD galaxies tend to be centered on the red
continuum images of the galaxies.

\item LSB galaxies in which we detected
 multiple HII regions have one very bright region with  others
 much fainter, mostly located at the edges of galaxies.

\item On average, there are $\sim$2 HII regions per LSB galaxy. They
are smaller and fainter than those found in the BCDs of the Virgo Cluster.

\item The fractional coverage of a galaxy by HII regions is similar in LSBs
 and in BCDs, and so is the equivalent width of 
the line produced by individual HII regions. 

\item The HII regions we detected produce more than half of the
total H$\alpha$ line emission of an LSB galaxy, and in many cases
the entire galaxy line emission.

\item The strength of the H$\alpha$ emission
from an HII region is correlated  with the intensity of the red
continuum underneath the HII region in both (HSB and LSB) samples.

\end{enumerate}

Our findings do not indicate that there is a fundamental and basic
difference between LSB and HSB irregular galaxies in their star-forming
properties as reflected in their H$\alpha$ emission and its relation to
the underlying red continuum light, but support the possibility that 
these two types of objects represent similar phenomena, which differ 
only in the intensity of the star formation process.

\section*{Acknowledgments}
 EA is supported by a special grant from the Ministry of Science and the 
Arts to develop TAUVEX, a UV space imaging experiment. AH acknowledges 
support from the US-Israel Binational Science Foundation and travel
grants from
the Sackler Institute of Astronomy. NB is grateful for continued 
support of the Austrian Friends of Tel Aviv University.  The Telescope
Allocation Committee of the Russian Academy of Sciences 6.0-m BTA 
is thanked for generous
allocation of telescope time over the years. Astronomical 
research at Tel Aviv University
is partly supported by a Center of Excellence award from the Israel  
Science Foundation. We acknowledge   Bruno Binggeli
for an updated catalog of the Virgo Cluster, G. Lyle Hoffman for additional
HI information on Virgo galaxies, and Liese Van Zee for copies of images
from her samples of galaxies. Rebecca Koopman gracefully set up
electronic access to her thesis, which furnished comparison images of
spiral galaxies.
We are grateful for constructive remarks from Sara Beck,  Lyle Hoffman,
Simon Pustilnik, Gotthard Richter, Liese van Zee, and
an anonymous referee.
This research made use of the NASA/IPAC Extragalactic Database (NED)
which is operated by the Jet Propulsion Laboratory, California Institute of
Technology, under contract with the National Aeronautics and Space
Administration.

\section*{References}

\begin{description}

\item Almoznino, E. 1996, PhD thesis, Tel Aviv University.

\item Almoznino, E. \& Brosch, N. 1998, MNRAS, in press (AB98).

\item Bessel, M.S. 1979, PASP, 91, 589.


\item Binggeli,  B., Sandage, A. \& Tammann, G.A. 1985, AJ,  90, 
1681 (VCC).

\item Brosch, N. 1992, QJRAS, 33, 27.

\item Brosch, N., Heller, A. \& Almoznino, E. 1998a, MNRAS, in press.

\item Brosch, N., Heller, A. \& Almoznino, E. 1998b, ApJ, 504, in press.

 \item de Blok W.J.G., van der Hulst J.M. \& Bothun, G.D. 1995,
 MNRAS, 274, 235.

\item de Boer, K.S., Braun, J.M., Vallenari, A. \& Mebold, U.
1998, A\&A, 329, L49.

\item Draper, N.R. \& Smith, H. 1981, {\it Applied Regression
Analysis}, New York: Wiley, p. 32.

\item Federspiel, M., Tammann, G.A. \& Sandage, A. 1998, ApJ, 495, 115.

\item Fouqu\'{e}, P., Bottinelli, L., Gouguenheim, L. \& Paturel, G. 1990,
ApJ, 349, 1.


\item Gallagher, J.S. \& Hunter, D.A. 1989, AJ,  98, 806.

\item Gavazzi, G., Catinella, B., Carrasco, L., Boselli, A. \&
Contursi, A. 1998, AJ, 115, 1745.

\item Gonzalez, A.H. \& Faber, S.M. 1997, ApJ, 485, 80.

\item Hodge, P. 1969, ApJ, 156, 847.

\item  Hoffman, G.L., Helou, G., Salpeter, E.E., Glosson, J. \& Sandage, A.
  1987, ApJS, 63, 247.


\item Hoffman, G.L., Williams, H.L., Salpeter, E.E., Sandage, A., Binggeli,
 B.  1989, ApJS, 71, 701.

\item Hoffman, G.L., Helou, G., Salpeter, E.E., Lewis, B. M.
  1989, ApJ,  339, 812.


\item Hunter, D.A. \& Gallagher, J.S. 1986, PASP, 98, 5.

\item Kennicutt, R.C., Tamblyn, P. \& Congdon, C.W. 1994, ApJ, 435, 22.


\item Marlowe, A.T., Meurer, G.R. \& Heckman, T.M 1997, ApJS, 112, 285.

\item McGaugh, S.S., Bothun G.D. \& Schombert J.M. 1995a, AJ, 110, 573.

\item McGaugh, S.S., Schombert, J.M. \&  Bothun G.D. 1995b, AJ, 109, 2019.



\item Skillman, E.D., Kennicutt, R.C. \& Hodge, P.W. 1989, ApJ, 347, 883.

\item van den Bergh, S. 1996, PASP, 108, 1091.



\end{description}

\newpage
{\small
\begin{deluxetable}{ccccccccc}
\tablecaption{LSB dwarf sample}
\small
\tablehead{\colhead{VCC} & \colhead{m$_B$} & \colhead{m$_{67}$}  
 & \colhead{a} & \colhead{e} & \colhead{f} &
\colhead{F(H$\alpha$)}
& \colhead{EW(H$\alpha$)} & \colhead{log(SFR)}
}

\startdata
17  & 15.20 & 15.95$\pm$0.06 &  46 & 5 & 24 & 14.78$\pm$0.68 & 105$\pm$15 & --1.36\nl
168 & 17.10 & 16.03$\pm$0.07 &  22 & 3 & 0 & -2.94$\pm$1.20  & 0 & 0 \nl
169 & 16.50 & 16.78$\pm$0.10 &  20 & 3 & 6 & 0.44$\pm$0.29  & 7$\pm$5 & --3.49 \nl
217 & 15.50 & 16.60$\pm$0.06 &  56 & 7 & 38 & 2.26$\pm$0.15  & 29$\pm$3 & --2.17 \nl
260 & 15.70 & 16.49$\pm$0.06 &  27 & 2 & 9 & 0.28$\pm$0.11  & 3$\pm$1 & --3.09 \nl
328 & 16.90 & 16.24$\pm$0.06 &  33 & 6 & 45 & 2.12$\pm$0.10  & 20$\pm$1 & --2.21 \nl
329 & 16.80 & 18.00$\pm$0.10 &   8 & 0 & 0 & 0.37$\pm$0.12  & 17$\pm$7 & --2.96 \nl
350 & 17.05 & 17.97$\pm$0.07 &  19 & 7 & 30 & 0.34$\pm$0.04  & 16$\pm$2 & --3.00 \nl
367 & 17.20 & 16.84$\pm$0.06 &   22 & 2 & 0 & -0.03$\pm$0.66 & 0 & 0 \nl
381 & 16.50 & 16.42$\pm$0.27 &  (24) & (7) & 0 & 0.16$\pm$0.43 & 2$\pm$5 & --3.33 \nl
477 & 16.96 & 16.96$\pm$0.15 &  27 & 0 & 6 & 0.33$\pm$0.14 & 7$\pm$3 & --2.96 \nl
530 & 15.80 & 15.43$\pm$0.12  &  41 & 2 & 6 & 0.77$\pm$0.31 & 3$\pm$1 &  --2.65 \nl
565 & 15.70 & 16.46$\pm$0.08 &   19 & 2 & 0 & 1.67$\pm$0.80 & 19$\pm$13 & --2.31 \nl
584 & 15.80 & 15.92$\pm$0.05  &  28 & 1 & 0 & 0.08$\pm$0.50 & 1$\pm$4 & --3.63 \nl
826 & 15.00 & 15.23$\pm$0.05 &   59 & 4 & 22 & 2.59$\pm$0.29 &  9$\pm$2 & --2.12 \nl
963  & 17.20 & 17.00$\pm$0.05 &   37 & 1 & 85 & 1.73$\pm$0.16 & 28$\pm$5 & --2.30 \nl
1455 & 16.80 & 16.29$\pm$0.07 &   23 & 4 & 75 & 1.70$\pm$0.13 & 17$\pm$2 & --2.30 \nl
1465 & 15.00 & 15.88$\pm$0.07 &   25 & 4 & 49 & 4.39$\pm$0.20 & 29$\pm$2 & --1.89 \nl
1468 & 15.00 & 15.87$\pm$0.09 &  28 & 4 & 25 & 3.50$\pm$0.31 & 23$\pm$3 & --1.99 \nl
1585 & 15.45 & 14.85$\pm$0.10 &  41 & 5 & 44 & 8.22$\pm$0.76 & 21$\pm$3 & --1.62 \nl
1753 & 16.81 & 18.04$\pm$0.06 &   16 & 2 & 81 & 0.68$\pm$0.05 & 32$\pm$4 & --2.70 \nl
1784 & 15.84 & 16.10$\pm$0.10 &   26 & 6 & 0 & 0.23$\pm$0.21 & 2$\pm$2 & --3.17 \nl
1816 & 16.20 & 16.16$\pm$0.15 &  28 & 6 & 0 & 1.87$\pm$0.60 & 16$\pm$7 & --2.26 \nl
1822 & 15.60 & 15.97$\pm$0.22 &   36 & 6 & 0 & -0.09$\pm$0.30 & 0 & 0 \nl
1952 & 16.00 & 16.63$\pm$0.10 &  22 & 5 & 45 & 2.39$\pm$0.70  & 32$\pm$16 & --2.15 \nl
1992 & 15.50 & 15.32$\pm$0.11 &  26 & 2 & 34 & 6.12$\pm$0.40 & 24$\pm$3 & --1.75 \nl
2034 & 15.82 & INDEF    &         26 & - & 0 & 1.54$\pm$0.50 & 2$\pm$1 & --2.35 \nl

\enddata

\tablecomments{The total blue magnitude is the estimate from the VCC
ande m$_{67}$ is the monochromatic magnitude of the continuum near 6700\AA\,. 
The blue magnitudes of VCC 1585, 1753, and 1784 are a private communication from
 Bruno Binggeli and are different from the values listed in the VCC.
 The semi-major axis a is in arcsec and is derived from an
ellipse fitting to the 25 mag arcsec$^{-2}$ isophote. The H$\alpha$ flux
is in units of 10$^{-14}$ erg cm$^{-2}$ s$^{-1}$. The ellipticity values 
are given as 10$\times$e and are rounded off to single digits. The
covering factor f is given in percent of the galaxy area in red continuum light. 
The SFR is in
M$_{\odot}$ yr$^{-1}$ and is obtained by assuming a uniform distance of
18 Mpc to all galaxies.}

\end{deluxetable}
}

\newpage
\newpage
{\small
\begin{deluxetable}{ccccccccc}
\tablecaption{HSB dwarf sample}
\small
\tablehead{\colhead{VCC} & \colhead{m$_B$} & \colhead{m$_R$}  
& \colhead{a} & \colhead{e} & \colhead{f} & \colhead{F(H$\alpha$)} 
& \colhead{EW(H$\alpha$)} & \colhead{log(SFR)} }
\startdata
10   & 16.20$\pm$0.09 & 14.88$\pm$0.10 & 24 & 5 & 98 & 9.30$\pm$1.90 & 31$\pm$7 & --1.56\nl
22   & 16.83$\pm$0.03 & 15.82$\pm$0.05 &  5 & 2 & 0  & 0.5$\pm$2.0 & 0 & 0 \nl
24   & 16.12$\pm$0.03 & 14.89$\pm$0.07 & 22 & 3 & 0  & 1.50$\pm$4.90 & 0 & 0 \nl
144  & 15.29$\pm$0.03 & 14.61$\pm$0.03 & 13 & 3 & 100 & 70.5$\pm$7.6 & 235$\pm$25 & --0.68\nl
172  & 15.38$\pm$0.03 & 14.53$\pm$0.03 & 27 & 6 & 0  & 0.37$\pm$0.60 & 1$\pm$2 & 0 \nl
324  & 14.58$\pm$0.03 & 13.62$\pm$0.02 & 29 & 2 & 74 & 63.1$\pm$4.3 & 210$\pm$14 & --0.73 \nl
410  & 17.76$\pm$0.04 & 16.89$\pm$0.04 &  9 & 2 & 100 & 0.50$\pm$0.05 & 1.7$\pm$0.2 & --2.83 \nl
459  & 15.29$\pm$0.04 & 14.40$\pm$0.07 & 22 & 2 & 38 & 29.7$\pm$3.3 & 99$\pm$11 & --1.06 \nl
513  & 15.84$\pm$0.03 & 14.56$\pm$0.02 & 13 & 2 & 52 & 21.7$\pm$2.0 & 72$\pm$7 & --1.20 \nl
562  & 16.59$\pm$0.03 & 15.86$\pm$0.05 & 11 & 3 & 48 & 13.4$\pm$0.9 & 45$\pm$3 & --1.41 \nl
985  & 16.62$\pm$0.04 & 15.61$\pm$0.07 & 13 & 5 & 23 & 3.1$\pm$1.8  & 10$\pm$6 & --2.04 \nl
1179 & 15.98$\pm$0.05 & 15.06$\pm$0.04 & 18 & 5 & 36 & 6.0$\pm$2.4  & 20$\pm$8 & --1.75 \nl
1374 & 15.00$\pm$0.02 & 14.19$\pm$0.02 & 32 & 6 & 59 & 29.7$\pm$1.3 & 99$\pm$4 & --1.06 \nl
1725 & 15.18$\pm$0.04 & 14.48$\pm$0.03 & 27 & 6 & 30 & 24.2$\pm$2.9 & 81$\pm$10 & --1.14 \nl
1791 & 14.85$\pm$0.04 & 14.09$\pm$0.03 & 36 & 7 & 29 & 40.3$\pm$1.9 & 134$\pm$6 & --0.93 \nl
2033 & 15.61$\pm$0.04 & 14.54$\pm$0.03 & 8 & 1 & 0  & 2.1$\pm$1.5  & 7$\pm$5   & --2.21 \nl
\enddata

\tablecomments{The total blue magnitude is the estimate from the VCC. 
 The semi-major axis a is in arcsec and is derived from an
ellipse fitting to the 25 mag arcsec$^{-2}$ isophote. The H$\alpha$ flux
is in units of 10$^{-14}$ erg cm$^{-2}$ s$^{-1}$. The ellipticity values 
are given as 10$\times$e and are rounded off to single digits. The
covering factor f is given in percent of the galaxy area in red continuum light.
The SFR is in
M$_{\odot}$ yr$^{-1}$ and is obtained by assuming a uniform distance of
18 Mpc to all galaxies.}

\end{deluxetable}
}

\newpage

\begin{deluxetable}{ccc}
\tablecaption{HII region statistics in LSB and HSB samples}
\small
\tablehead{\colhead{Parameter } & \colhead{ BCD} &
 \colhead{LSB }  }
\startdata
 HII regions present   &  12/16 (75\%)           & 17/27 (63\%)  \nl
                       & [8/12 (67\%) pure BCD]  &               \nl
 \hline
Single HII region        & 9/12 (75\%)             & 7/17 (42\%)   \nl
\hline
Multiple HII regions  & 3/12 (25\%)            &  10/17 (49\%) \nl
 \hline
Covering factor of    & 67\% (pure BCD)          & 37\%         \nl
``continuum'' galaxy    & 57\% (all galaxies)             &             \nl
\hline
 \enddata

\tablecomments{The covering factor listed in the last row is obtained by 
dividing the summed area of all HII regions by the area of the galaxy, as defined 
by the 25 mag arcsec$^{-2}$ isophote on the red continuum image.}

\end{deluxetable}

\newpage

\begin{deluxetable}{cccccc}
\tablecaption{Properties of individual HII regions in LSB galaxies}
\small
\tablehead{\colhead{Galaxy (VCC)} & \colhead{HII region ID} &
 \colhead{F(H$\alpha$)} & \colhead{f(OFF)} & \colhead{EW(H$\alpha$)} & 
\colhead{log(SFR)}}
\startdata
17   & a & 6.92$\pm$0.15 & 15$\pm$1.4 & 461 & --1.70 \nl
17   & b & 0.46$\pm$0.04 & 6.0$\pm$0.7 & 77 & --2.87 \nl
17   & c & 0.54$\pm$0.05 & 7.0$\pm$0.8 & 77 & --2.80 \nl
17   & d & 0.68$\pm$0.05 & 6.0$\pm$0.7  & 113 & --2.70 \nl
17   & e & 0.48$\pm$0.04 & 2.0$\pm$0.5  & 240 & --2.85 \nl
17   & f & 1.49$\pm$0.09 & 3.0$\pm$2.0  & 497 & --2.37 \nl
169  & a & 0.37$\pm$0.04 & 0.8$\pm$0.4& 462 & --2.96 \nl
169  & b & 0.15$\pm$0.03 & 4.0$\pm$0.7   & 38 & --3.36 \nl
169  & c & 0.16$\pm$0.03 & 0.8$\pm$0.5  & 200 & --3.33 \nl
217  & a & 0.73$\pm$0.06 & 4.0$\pm$0.1  & 183 & --2.67 \nl
217  & b & 0.73$\pm$0.06 & 10$\pm$1   & 73 & --2.67 \nl
217  & c & 0.20$\pm$0.03 & 4.0$\pm$0.8  & 50 & --3.23 \nl
260  & a & 0.18$\pm$0.03 & 2.7$\pm$0.5 & 67 & --3.29 \nl
328  & a & 2.15$\pm$0.14 & 114$\pm$3   & 19 & --2.20 \nl
350  & a & 0.22$\pm$0.05 & 3.0$\pm$0.6  & 73 & --3.19 \nl
477  & a & 0.27$\pm$0.04 & 5.0$\pm$0.8  & 54 & --3.10 \nl
530  & a & 0.26$\pm$0.03 & 19$\pm$1   & 14 & --3.12 \nl
530  & b & 0.18$\pm$0.03 & 18$\pm$1   & 10 & --3.28 \nl
826  & a & 1.79$\pm$0.06 & 60$\pm$2    & 30 & --2.28 \nl
826  & b & 0.11$\pm$0.02 & 30$\pm$1    & 4  & --3.49 \nl
826  & c & 0.40$\pm$0.03 & 10$\pm$30     & 40 & --2.93 \nl
826 & d & 0.31$\pm$0.03 & 3.0$\pm$0.3   & 103 & --3.04 \nl
963  & a & 1.65$\pm$0.08 & 51$\pm$2  & 32 & --2.32 \nl
1455 & a & 1.65$\pm$0.09 &  90$\pm$3   & 18 & --2.31\nl
1465 & a & 3.04$\pm$0.06 & 50$\pm$1    & 61 & --2.05 \nl
1465 & b & 0.99$\pm$0.04 & 30$\pm$1    & 33 & --2.53 \nl
1468 & a & 3.17$\pm$0.08 & 36$\pm$1   & 88 & --2.03 \nl
1585 & a & 3.94$\pm$0.14 & 76$\pm$3   & 52 & --1.93 \nl
1585 & b & 1.74$\pm$0.10 & 74$\pm$3   & 24 & --2.29 \nl
1585 & c & 0.49$\pm$0.07 & 50$\pm$3    & 10 & --2.84 \nl
1585 & d & 0.38$\pm$0.05 & 10$\pm$1    & 38 & --2.95 \nl
1753 & a & 0.32$\pm$0.06 & 10$\pm$1   & 32 & --3.02 \nl
1753 & b & 0.20$\pm$0.04 & 10$\pm$1   & 20 & --3.23 \nl
1753 & c & 0.13$\pm$0.03 & 5$\pm$1   & 26 & --3.42 \nl
1952 & a & 1.31$\pm$0.09 & 30$\pm$2    & 44 & --2.41 \nl
1952 & b & 0.80$\pm$0.08 & 70$\pm$3    & 11 & --2.63 \nl
1992 & a & 2.45$\pm$0.10 & 40$\pm$2    & 61 & --2.14 \nl
1992 & b & 2.53$\pm$0.11 & 90$\pm$3    & 28 & --2.13 \nl
\enddata

\tablecomments{F(H$\alpha$) is in 10$^{-14}$ erg cm$^{-2}$ s$^{-1}$ and
f(OFF) is in 10$^{-17}$ erg cm$^{-2}$ s$^{-1}$ \AA\,$^{-1}$.
The SFR calculation is explained in the text. The equivalent width of the line
is obtained by EW=$\frac{F(H\alpha)}{f(OFF)}$. The HII region labelled
$f$ in VCC 17 appears on a particularly faint part of the galaxy; we could
not measure the underlying continuum accurately enough to 
provide a good quality H$\alpha$ equivalent width.}

\end{deluxetable}
 
\newpage

\newpage

\begin{deluxetable}{cccccc}
\tablecaption{Properties of individual HII regions in HSB galaxies}
\small
\tablehead{\colhead{Galaxy (VCC)} & \colhead{HII region ID} &
 \colhead{F(H$\alpha$)} & \colhead{f(OFF)} & \colhead{EW(H$\alpha$)} & 
\colhead{log(SFR)}}
\startdata
10   & a & 8.57$\pm$1.76        & 0.23$\pm$0.05 & 37   &	  --1.60 \nl
144  & a & 71.00$\pm$3.23	& 0.42$\pm$0.07	& 169  &    	  --0.68 \nl
324  & a & 60.9$\pm$4.21	& 0.59$\pm$0.10	& 104  &    	  --0.75 \nl
410  & a & 4.16$\pm$0.38	& 0.05$\pm$0.01  & 90   &	  --1.91 \nl
459  & a & 20.5$\pm$2.36	& 0.23$\pm$0.04	& 87   &	  --1.22 \nl
513  & a & 20.7$\pm$1.91	& 0.31$\pm$0.06	& 68   &	  --1.22 \nl
562  & a & 12.7$\pm$0.88	& 0.08$\pm$0.01  & 154  &	  --1.43 \nl
985  & a & 1.44$\pm$0.23	& 0.05$\pm$0.02	& 30   &	  --2.37 \nl
1179 & a & 5.77$\pm$2.26	& 0.13$\pm$0.05 & 46   &	  --1.77 \nl
1374 & a & 21.3$\pm$0.98	& 0.25$\pm$0.05	& 84   &	  --1.20 \nl
1374 & b & 3.86$\pm$0.18	& 0.18$\pm$0.08	& 22   &	  --1.95 \nl
1374 & c & 1.78$\pm$0.08	& 0.06$\pm$0.06	& 32   &	  --2.29 \nl
1725 & a & 10.13$\pm$1.17	& 0.10$\pm$0.02	& 98   &	  --1.53 \nl
1725 & b & 5.86$\pm$0.68	& 0.09$\pm$0.02	& 63   &	  --1.76 \nl
1791 & a & 7.25$\pm$0.33	& 0.05$\pm$0.01 & 135  &	  --1.67 \nl
1791 & b & 5.89$\pm$0.27	& 0.05$\pm$0.01 & 129  &	  --1.76 \nl
1791 & c & 11.1$\pm$0.51	& 0.15$\pm$0.03	& 76   &	  --1.49 \nl
1791 & d & 2.19$\pm$0.10	& 0.07$\pm$0.03	& 30   &	  --2.19 \nl
\enddata

\tablecomments{F(H$\alpha$) and f(OFF) are in 10$^{-14}$ erg cm$^{-2}$ s$^{-1}$.
The SFR calculation is explained in the text. The equivalent width of the line
is obtained by EW=$\frac{F(H\alpha)}{f(OFF)}$.}

\end{deluxetable}

\section*{Figure captions}
\begin{itemize}

\item Fig. 1: Distribution of LSB dwarfs studied here and HSB dwarfs
from AB98 projected on the Virgo cluster. The different sub-clusterings
identified by Hoffman {\it et al.} are indicated schematically. 
BCD galaxies are represented
as filled circles and LSBs are empty circles. The size of the symbols 
is proportional to the blue magnitude of each galaxy in the VCC.

\item Fig. 2: Mosaic of H$\alpha$ line (left), continuum (center), and 
net-H$\alpha$ (right) images
of four galaxies in the LSB sample. North is up and East is to the left.
The scale is indicated by a 10 arcsec long bar 
in the top right corner. The galaxies shown are (top to bottom)
VCC 17, VCC 169, VCC 217, and VCC 260.

\item Fig. 3:  Same as Fig. 2. The galaxies shown are VCC 328, VCC 329, VCC 350, and
VCC 367.

\item Fig. 4:  Same as Fig. 2. The galaxies shown are VCC 381, VCC 477, VCC 530,
and VCC 565.

\item Fig. 5:  Same as Fig. 2. The galaxies shown are VCC 584, VCC 826, VCC 1455,
and VCC 1465.

\item Fig. 6:  Same as Fig. 2. The galaxies shown are VCC 1585, VCC 1753, VCC 1784,
and VCC 1816.

\item Fig. 7:  Same as Fig. 2. The galaxies shown are VCC 1822, VCC 1952, VCC 1992, 
and VCC 2034.

\item Fig. 8:   Same as Fig. 2. The galaxies shown are VCC 1822, VCC 1952, VCC 1992, 
and VCC 2034.

\item Fig. 9:  Mosaic of line (left), continuum (center), and net-H$\alpha$ (right) images
of four galaxies in the HSB sample. North is up and East is to the left.
The scale is indicated by a 10 arcsec long bar 
in the top right corner for each object. The galaxies shown are (top to bottom)
VCC 10, VCC 144, VCC 172, and VCC 324.

\item Fig. 10: Same as Fig. 9. The galaxies shown are VCC 410, VCC 459, 
VCC 513, VCC 562, and VCC 985.

\item Fig. 11: Same as Fig. 9. The galaxies shown are VCC 1179, VCC 1374, VCC 1725, 
and VCC 1791.


\item Fig. 12: Distribution of star formation rates for individual HII regions 
in the HSB and LSB galaxy samples. 

\item Fig. 13: Correlation between the logarithms of the H$\alpha$ emission 
and of the red continuum light
for individual HII regions in LSB galaxies.  The data for HII regions of
BCDs is plotted as squares and that for the HII regions
of LSBs is represented by stars. The units are 10$^{-14}$ erg cm$^{-2}$ s$^{-1}$
for the line flux and 10$^{-14}$ erg cm$^{-2}$ s$^{-1}$ \AA\,$^{-1}$ for
the continuum flux density.

\item Fig. 14:  Distribution of the coverage factor of a galaxy by HII regions. 

\item Fig. 15: H$\alpha$ line contrast {\it vs.} off-line continuum
contrast between the brightest and the faintest HII regions we detected in
galaxies with multiple HII regions.

\end{itemize}

\newpage
   
\begin{figure}
\vspace{17cm}
\includegraphics{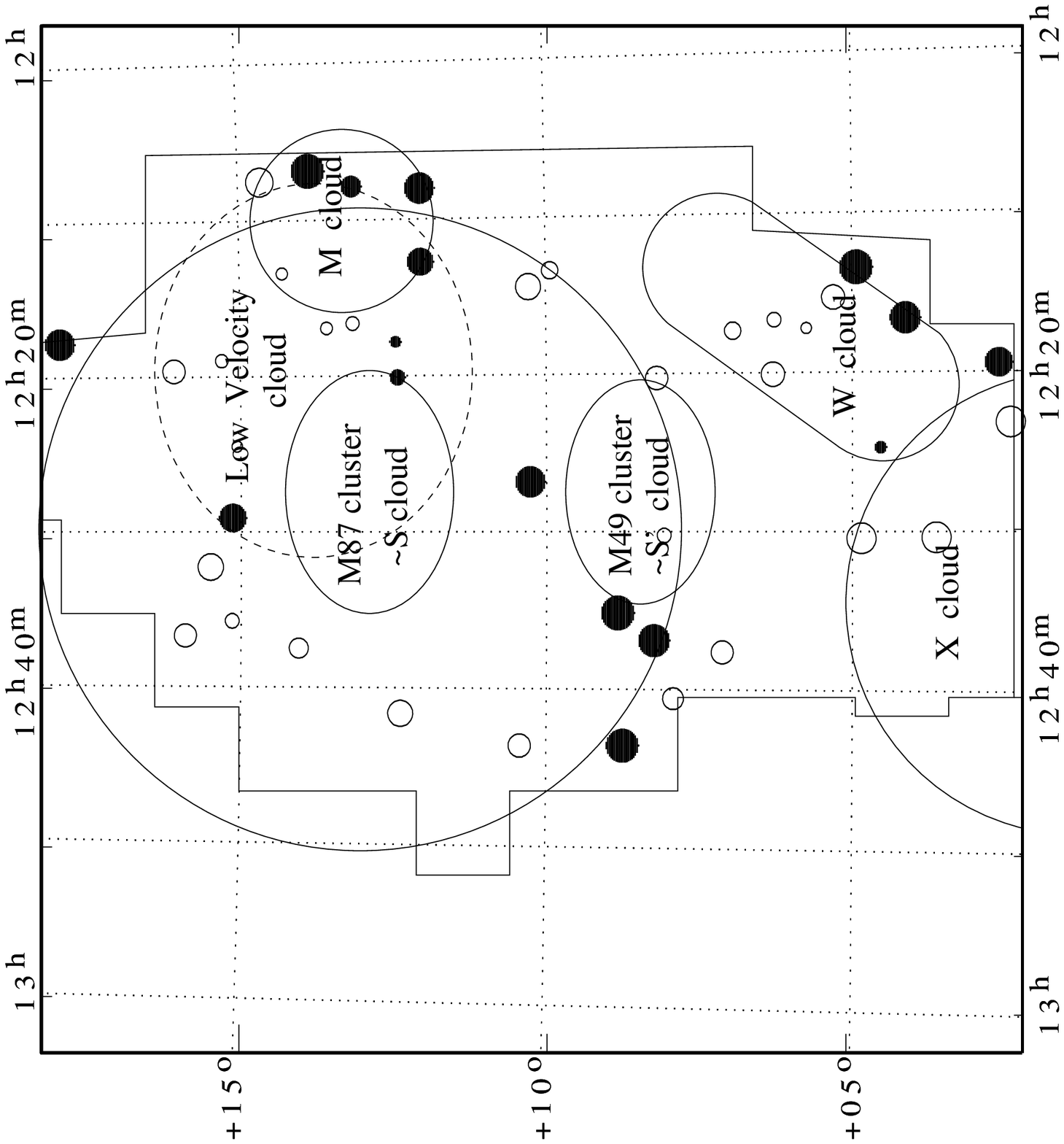}
\end{figure} 

   \newpage

\begin{figure}
\vspace{15cm}
\includegraphics{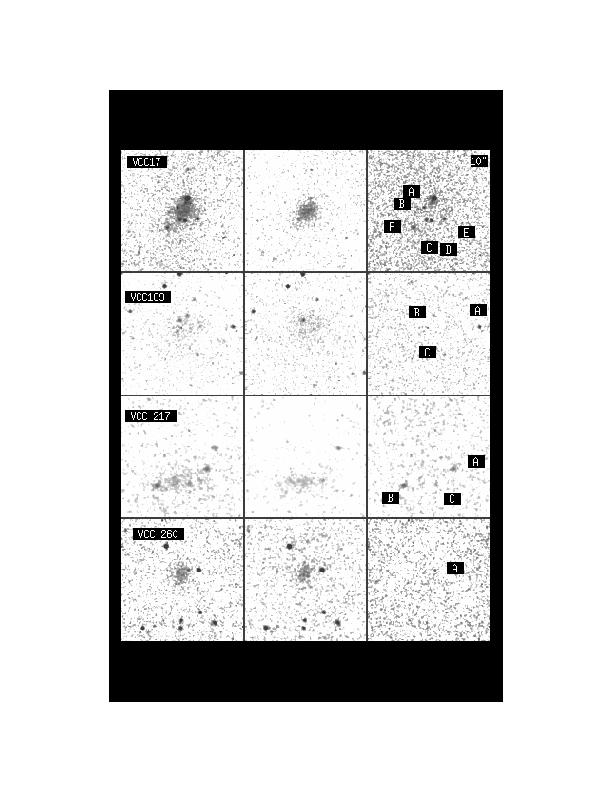}
\end{figure} 

    \newpage

\begin{figure}
\vspace{15cm}
\includegraphics{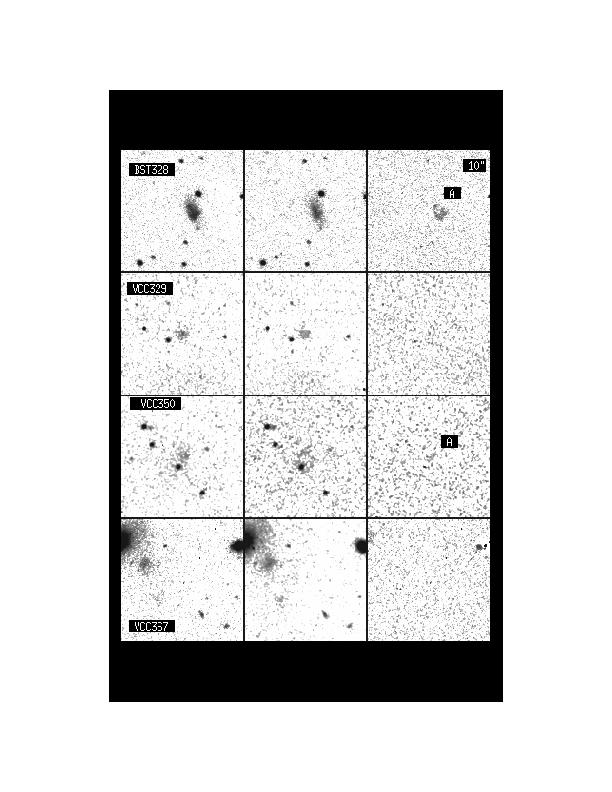}
\end{figure} 

 \newpage
   
\begin{figure}
\vspace{15cm}
\includegraphics{mos4x3c.a.new.ps}
\end{figure} 

    \newpage

\begin{figure}
\vspace{15cm}
\includegraphics{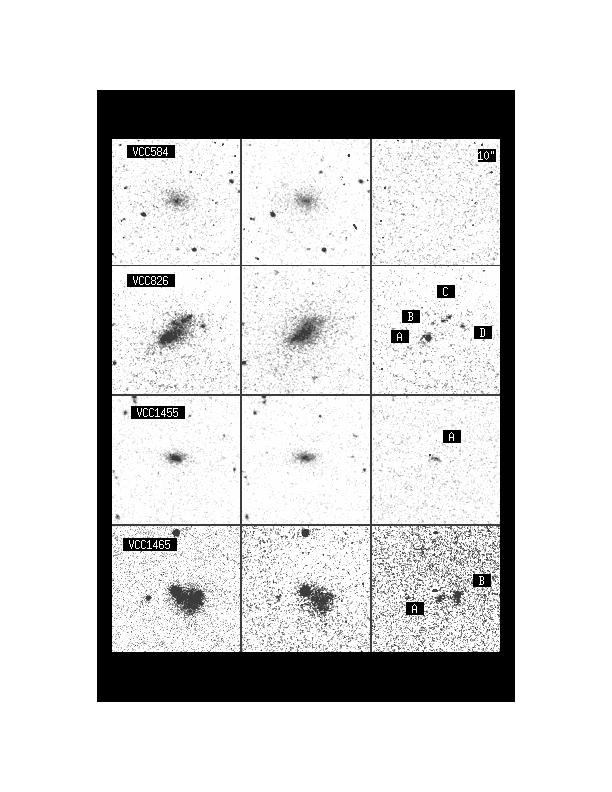}
\end{figure} 

 \newpage
   
\begin{figure}
\vspace{15cm}
\includegraphics{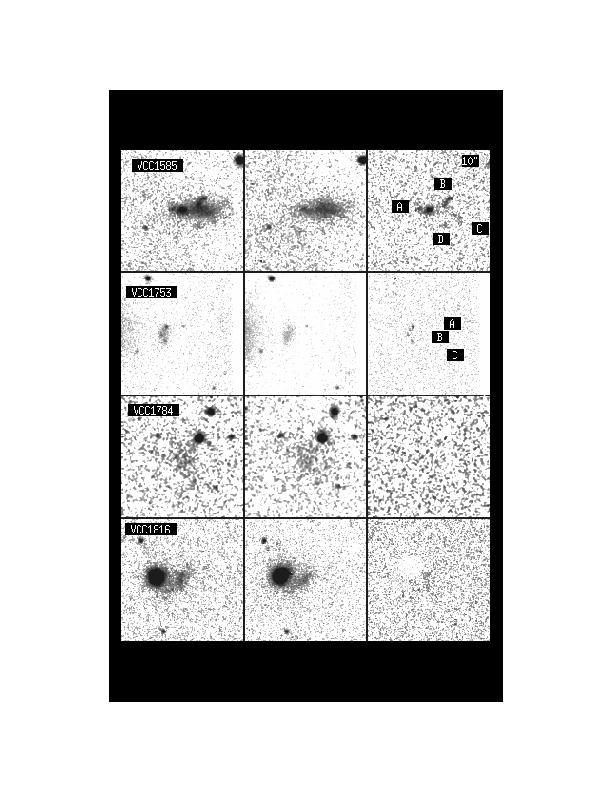}
\end{figure} 

    \newpage

\begin{figure}
\vspace{15cm}
\includegraphics{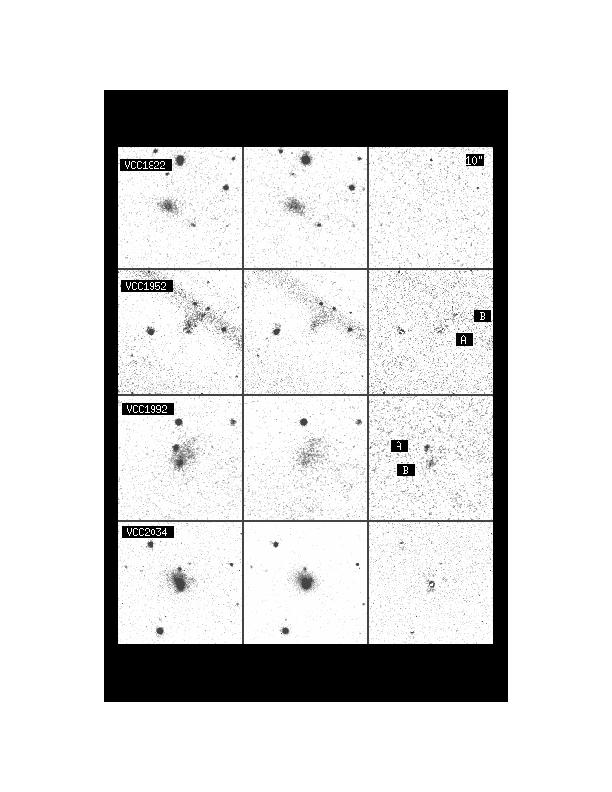}
\end{figure}

\newpage
   
\begin{figure}
\vspace{15cm}
\includegraphics{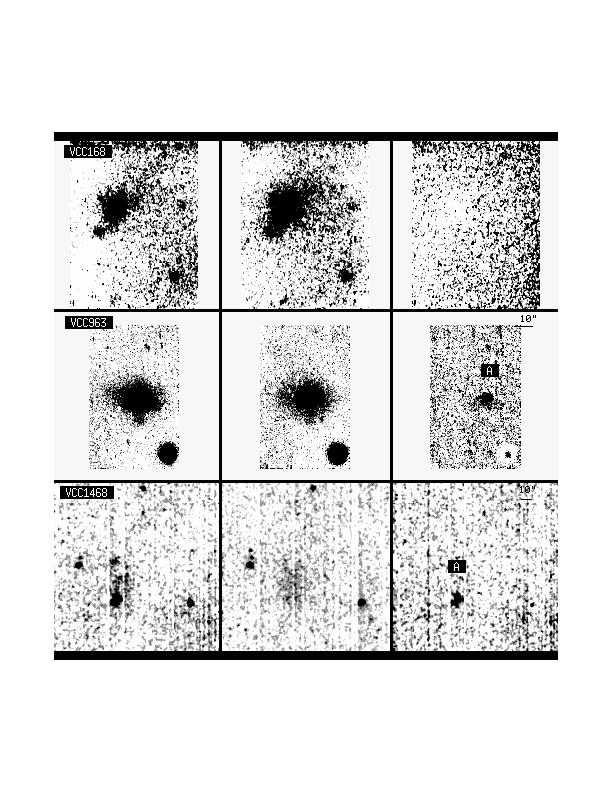}
\end{figure} 

  \newpage
   
\begin{figure}
\vspace{15cm}
\includegraphics{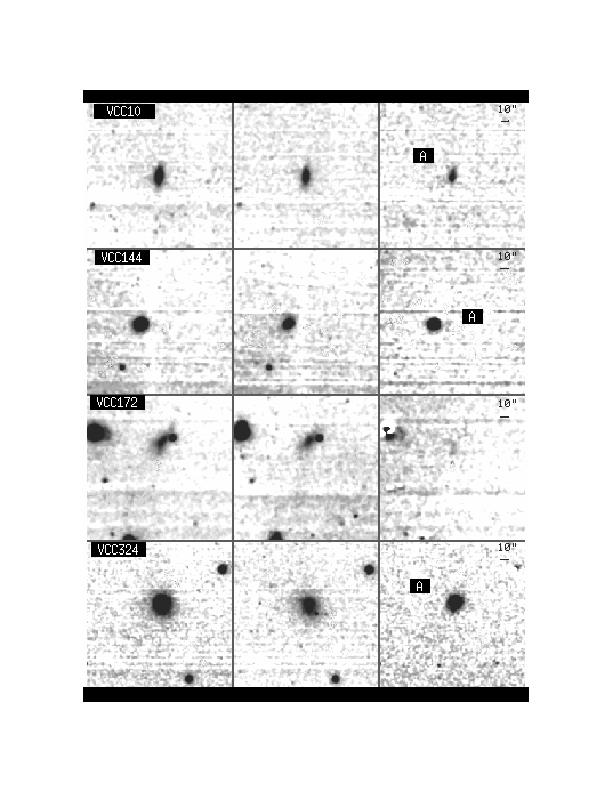}
\end{figure} 

\newpage

\begin{figure}
\vspace{15cm}
\includegraphics{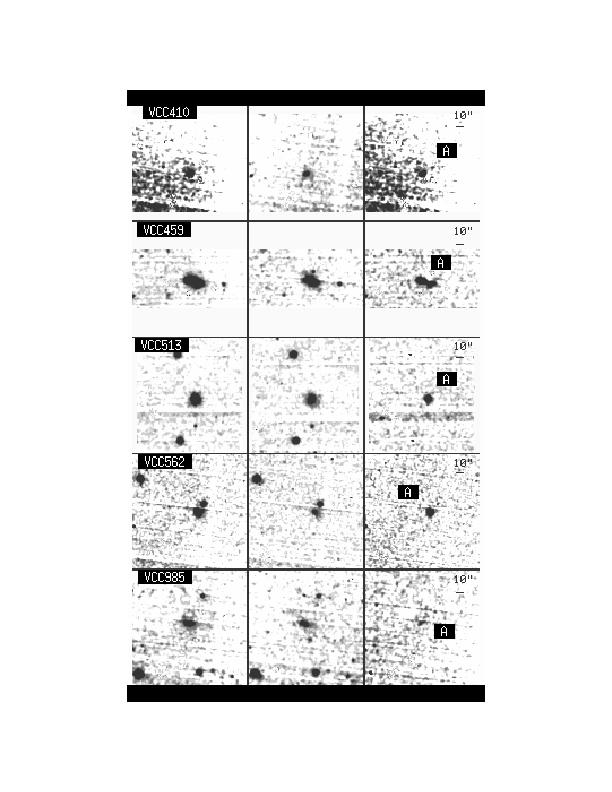}
\end{figure}

\newpage

\newpage
\begin{figure}
\vspace{15cm}
\includegraphics{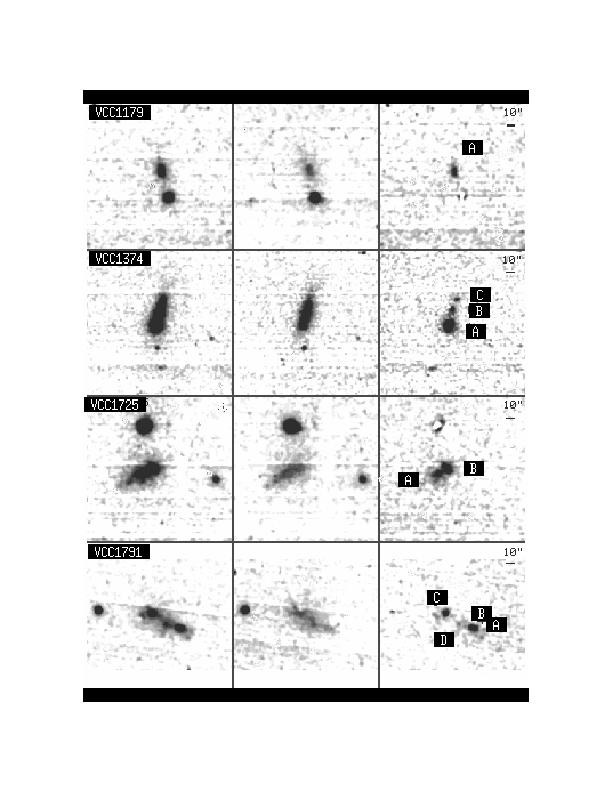}
\end{figure}

\newpage
 
\begin{figure} 
\vspace{15cm}
\putplot{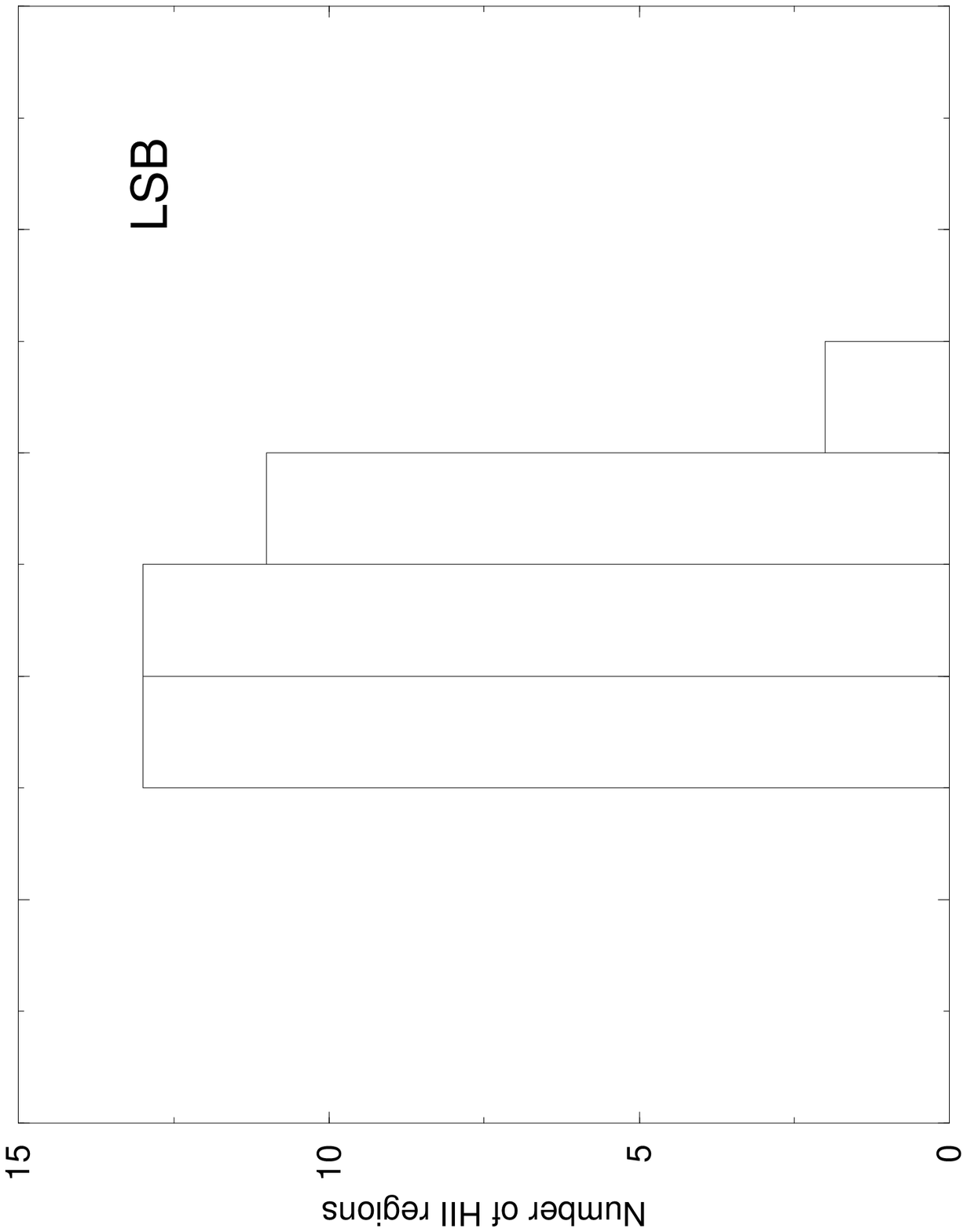}{1in}{-90}{60}{60}{-250}{446}
\putplot{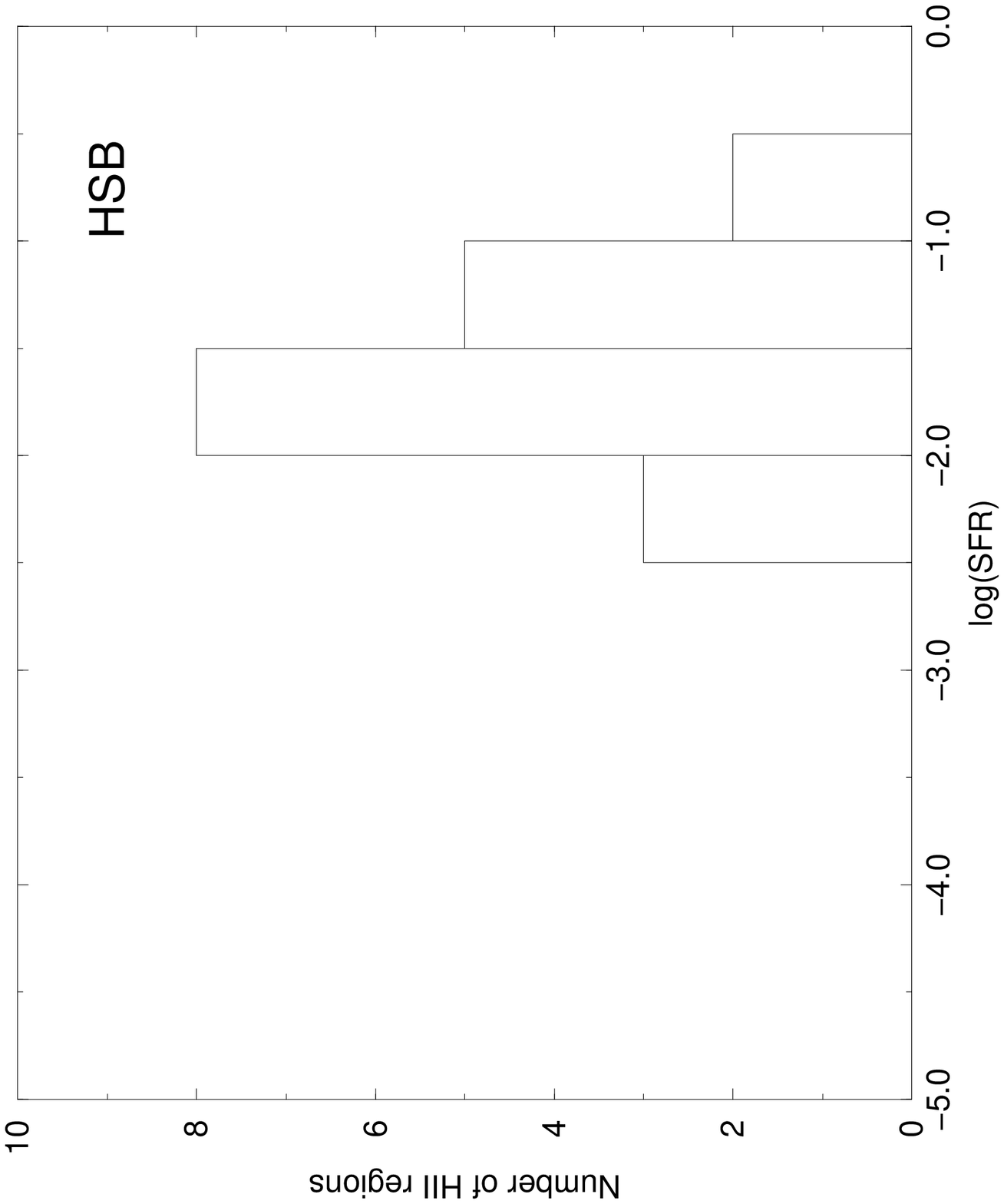}{1in}{-90}{60}{60}{-250}{300}
\end{figure}

 \newpage

\begin{figure}
\vspace{11cm}
\includegraphics{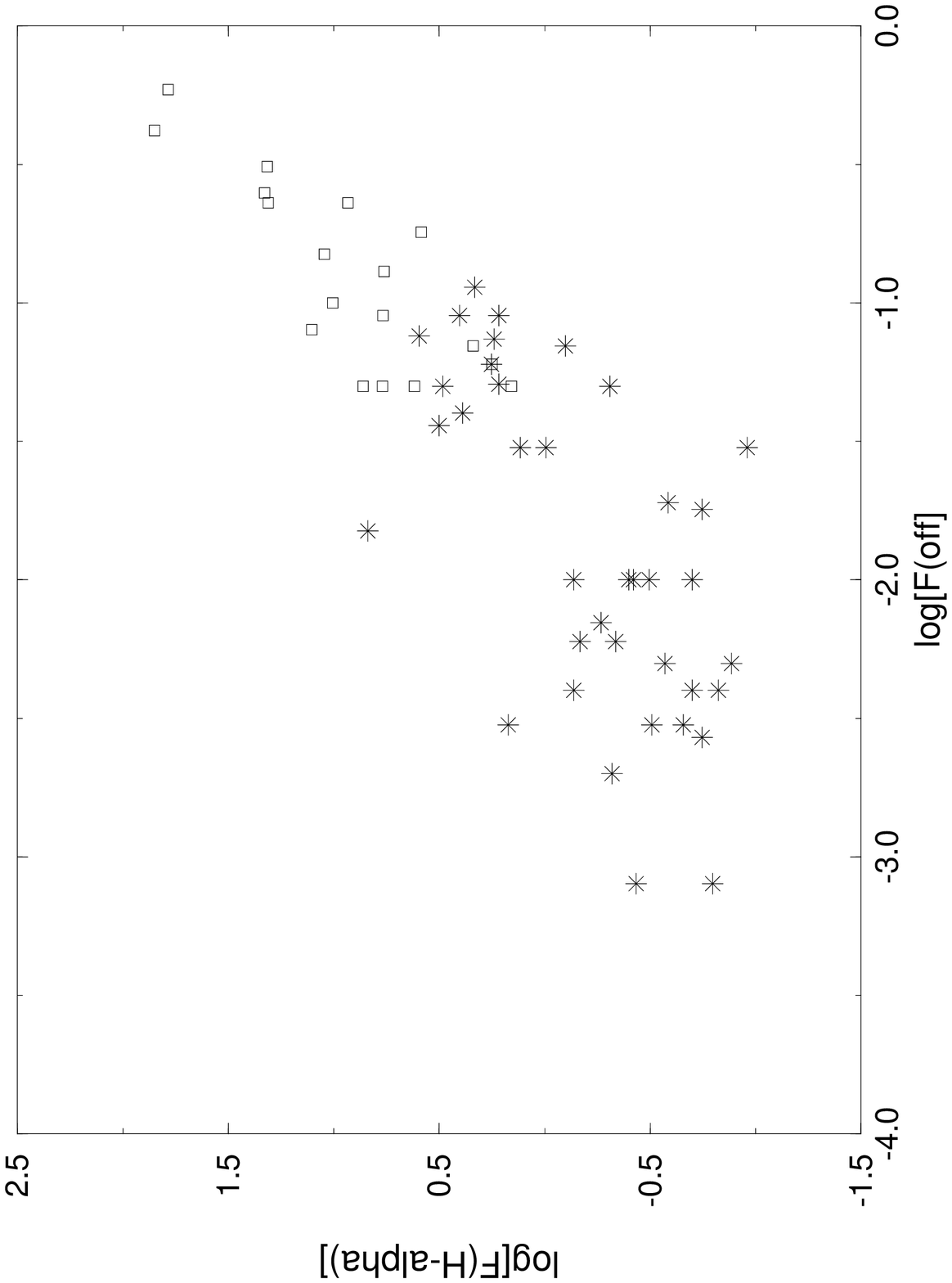}
\end{figure}

\newpage

\begin{figure}
\vspace{15cm}
\putplot{FfracLSB.eps}{1in}{-90}{60}{60}{-250}{446}
\putplot{FfracHSB.eps}{1in}{-90}{60}{60}{-250}{300}
 
\end{figure}
\pagebreak
 \newpage

\begin{figure}
\vspace{10cm}
\includegraphics{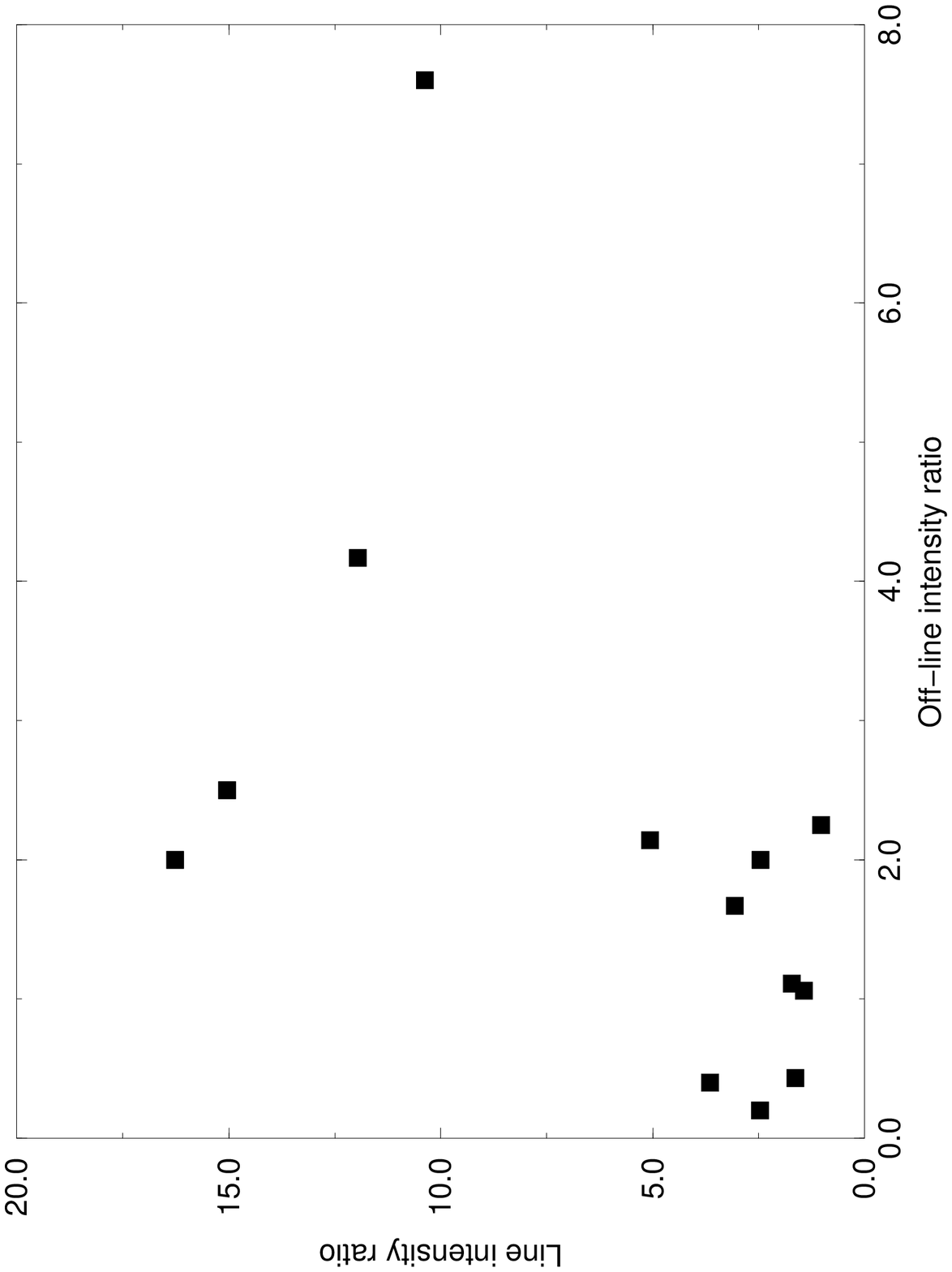}
\end{figure}

\end{document}